\documentclass[pdflatex,sn-basic,Numbered]{sn-jnl}

\usepackage{graphicx}
\usepackage{multirow}
\usepackage{amsmath,amssymb,amsfonts}
\usepackage{amsthm}
\usepackage{mathrsfs}
\usepackage[title]{appendix}
\usepackage{xcolor}
\usepackage{textcomp}
\usepackage{manyfoot}
\usepackage{booktabs}
\usepackage{listings}
\usepackage{tikz}

\usepackage{tabularx}
\usepackage{array}
\usepackage{ragged2e}

\newcolumntype{L}[1]{>{\RaggedRight\arraybackslash}p{#1}}
\newcolumntype{Y}{>{\RaggedRight\arraybackslash}X}

\usepackage{longtable}
\usepackage{array}
\usepackage{colortbl}

\newcolumntype{L}[1]{>{\RaggedRight\arraybackslash}p{#1}}

\newlength{\fslength}
\makeatletter

\newcommand{\funnyP}{%
    \setlength{\fslength}{\f@size pt}%
    \reflectbox{\ensuremath{\mathrm{P}}}%
    \hspace*{-.300\fslength}%
    \ensuremath{\mathrm{P}}%
}

\newcommand{\funnyPbar}{%
    \setlength{\fslength}{\f@size pt}%
    \reflectbox{\ensuremath{\bar{\mathrm{P}}}}%
    \hspace*{-.300\fslength}%
    \ensuremath{\bar{\mathrm{P}}}%
}

\makeatother

\begin{document}

\title[Productivity in HPC]{Productivity in HPC: A Critical Review from SLOC and Performance Portability to AI-Assisted Programming}

\author{\fnm{Ami} \sur{Marowka}}
\email{amimar2@yahoo.com}

\affil{\orgdiv{Parallel Research Lab},
           \orgaddress{\city{Petah-Tikva}, \country{Israel}}}

\abstract{Productivity has long been recognized as a central concern in
high-performance computing (HPC), yet it still lacks a workable formal
definition and a broadly accepted metric for quantified assessment.
Recent attempts to incorporate productivity into performance-portability
analysis have introduced code-divergence, convergence, and semantic
measures, but have not established a convincing formal relationship
between productivity and performance portability. 

\smallskip
More recent language and compiler systems, including JACC, Mojo, JLIR,
and Accl++, pursue productive performance portability through higher-level
abstractions, multi-level intermediate representations, automatic kernel
generation, and runtime compilation. These systems demonstrate substantial
engineering progress, while productivity is still commonly represented
through programmability, ecosystem reuse, code size, or reduced
implementation burden rather than direct measurement of human effort.

\smallskip
This paper presents a
critical review of HPC productivity measurement, tracing its development
from source lines of code (SLOC), static complexity, and empirical
usability studies to development-time models and semantic approaches.
We also examine how automated migration and artificial-intelligence(AI)-assisted programming
weaken the correspondence between code transformation and human effort,
shifting effort toward verification, correction, and performance tuning.
The reviewed methods are compared in terms of what they measure,
directness, objectivity, reproducibility, practicality, and their
relationship to performance and portability. We conclude by identifying
the principal limitations and open challenges for quantified HPC
productivity assessment.}


\maketitle


\section{Introduction}
\label{sec:introduction}

Performance has traditionally dominated the evaluation of high-performance
computing (HPC) systems and applications. Execution time, throughput, speedup,
efficiency, scalability, and related quantities provide relatively direct
observations of how effectively computational resources are used. Yet the value
of an HPC programming approach cannot be determined from performance alone.
The effort required to construct, parallelize, port, optimize, debug, and
maintain an implementation is also important, particularly as hardware
architectures and programming environments become increasingly heterogeneous.

For this reason, productivity has been studied in parallel and HPC programming
for several decades. Early studies frequently approached the problem through
\emph{programmability} or \emph{usability}. They examined how easily programmers
could learn a language, express parallelism, obtain a correct program, debug it,
and achieve an acceptable level of performance. Measures included programming
time, lines of code, debugging time, correctness, code expansion, subjective
assessments, and observations of programmer behavior
\cite{Hochstein2005,Ebcioglu2006,Marowka2013}. A recurring difficulty was
already apparent: productivity is influenced not only by a programming language
or programming model, but also by programmer experience, training, the problem
being implemented, development tools, hardware, and the performance target being
pursued. Marowka~\cite{Marowka2013}, for example, reviewed this heterogeneous
empirical literature and argued that usability directly affects productivity,
while also emphasizing that a programming model that is easy to use but fails
to provide adequate scalability cannot meaningfully be considered successful.

A second tradition approached productivity indirectly through properties of the
produced source code. Source Lines of Code (SLOC or LOC) and Logical Lines of Code (LLOC)
became particularly
common because they are inexpensive, automatic, and apparently objective.
Nguyen et al.~\cite{NguyenSLOC} note that SLOC became embedded in major
software-cost models partly because source code was historically the most
visible product of programming work. They also emphasize, however, that
inconsistent definitions and counting practices can make SLOC measurements
incomparable. In HPC, these difficulties are amplified because parallel
programming constructs of radically different conceptual complexity may occupy
the same number of lines.

Wienke et al.~\cite{Wienke2016} provide one of the most comprehensive
examinations of this problem in an HPC context. They review LOC, function
points, Halstead complexity, cyclomatic complexity, and development time, and
conclude that conventional software metrics either fail to capture important
parallel-programming activities or are poor predictors of actual development
effort. In particular, LOC assigns the same apparent cost to a simple
declaration and to a one-line OpenMP directive that may require substantial
reasoning about parallel correctness and performance. It also ignores work
performed outside the source text, such as compiler-option selection, process
binding, debugging, testing, and performance tuning.

The emergence of \emph{performance portability} added another dimension to the
productivity problem. Once applications were expected not merely to execute on
multiple architectures but to perform efficiently across them, productivity
increasingly became discussed together with performance and portability---the
so-called three Ps---performance, portability, and productivity---or P3. 
Harrell et al.~\cite{Harrell2018} explicitly argued
that evaluating performance-portable software should include both the efficiency
achieved and the cost of developing and maintaining the implementations. Their
study combines performance-portability measurements with development hours,
source changes, churn, code divergence, and subjective measures of difficulty.
Later, Pennycook et al.~\cite{Pennycook2021P3} transformed code divergence into
code convergence and proposed plotting it against performance portability in a
two-dimensional navigation chart. Lin, Deakin, and McIntosh-Smith
\cite{Lin2024Productivity} subsequently replaced purely textual divergence with
a tree-based approach designed to capture syntactic and semantic differences
between programming-model implementations.

A more recent line of work approaches the productivity problem from a
different direction: rather than defining another productivity measure,
it attempts to engineer programming and compiler systems that reduce the
amount of architecture-specific work required from the programmer.
JACC and JACC.Multi provide common high-level Julia interfaces over
heterogeneous CPU, GPU, and multi-GPU execution paths
\cite{ValeroLara2024JACC,ValeroLara2025JACCMulti}; Mojo uses MLIR-based
compilation together with Python interoperability to pursue portable GPU
programming~\cite{Godoy2025Mojo}; JLIR brings an MLIR-inspired
intermediate representation into Julia and automatically extracts portable
JACC kernels~\cite{Miniskar2026JLIR}; and Accl++ combines an
architecture-agnostic C++ interface with runtime compilation
\cite{Gonzalez2026Accl}. These systems represent an important shift from
\emph{measuring productivity} toward \emph{engineering for productivity}.
They therefore provide useful contemporary evidence for examining what
modern HPC studies mean when they claim improved productivity and how
that claim is actually evaluated.


A further challenge has emerged from the increasing automation of software
development. Source-to-source migration tools can transform substantial portions
of an implementation from one programming model to another. For example, tools
such as the Intel Data Parallel C++ (DPC++) Compatibility Tool and its open-source counterpart
SYCLomatic can automatically migrate CUDA
source code toward SYCL, a single-source C++ programming model for heterogeneous
computing. In such a
workflow, a large number of source lines may be added, removed, or modified
without requiring a corresponding amount of direct human programming effort.
The remaining effort may instead lie in resolving unsupported constructs,
verifying correctness, adapting the generated code, and recovering or improving
performance on the target platforms. Consequently, measures based on SLOC,
source changes, or code divergence may characterize the magnitude of the
resulting source transformation without necessarily characterizing the human
effort that produced it.

Generative artificial intelligence(AI) introduces a related but broader challenge. AI-assisted
programming can generate, translate, restructure, debug, and optimize source
code, further weakening the traditional correspondence between code quantity
and programmer effort. At the same time, the human contribution does not
disappear; rather, part of it may shift from direct code production toward
problem specification, evaluation of generated solutions, correctness
verification, performance analysis, and iterative refinement. Productivity
measurement in modern HPC must therefore address not only how much code is
produced or changed, but also which parts of the development process remain
meaningful indicators of human effort when substantial portions of the source
transformation are automated.

Recent empirical studies further show that conclusions about AI-induced
productivity gains can change substantially depending on whether productivity
is measured through code output, task completion time, completed work, or
other software-quality and developer-related outcomes
\cite{Paradis2025,MartinLopez2026,Gambacorta2026}.


These developments represent clear methodological progress, but they also expose
several unresolved questions. \emph{What exactly is being measured when an HPC
study reports ``productivity,'' and what relationship, if any, does that
quantity have to achieved performance, portability, or performance
portability?} A source-code size is not development effort; development time is
not independent of programmer experience; and code divergence is not
necessarily proportional to the cost of producing the divergence. Moreover,
when source code is generated or transformed automatically, the amount of code
created or modified may become even less representative of human effort.
Similarly, placing a productivity proxy beside a performance-portability metric
does not by itself establish a mathematical relationship between them. These
issues raise a broader question for modern HPC: which measurable quantities
remain meaningful indicators of productivity when programming, porting, and
optimization are increasingly assisted by automated tools?

This paper reviews the historical development of these approaches and makes four
main contributions. First, it provides a consolidated historical account of HPC
productivity measurement, ranging from empirical usability studies and SLOC to
development-effort models and semantic code-divergence metrics. Second, it
proposes a taxonomy that distinguishes what each method actually measures from
what it is intended to represent. 
Third, it examines separately the degree to which each approach is objective,
reproducible, intuitive, and practical for ordinary HPC studies, including
whether its interpretation remains meaningful under automated source migration
and AI-assisted programming.
Finally, it critically reviews attempts to connect
productivity with performance, portability, and formal performance-portability
metrics, distinguishing \emph{co-reporting and joint visualization} from an
actual quantitative relationship among the quantities.

For convenient reference, Appendix~\ref{app:glossary} provides a glossary of
the principal acronyms and technical terms used throughout the review.


\section{Scope and Review Method}
\label{sec:scope}

This work is a focused critical review rather than an exhaustive systematic
literature review. The historical corpus is constructed primarily by backward
citation tracing from two complementary anchor studies.

Marowka~\cite{Marowka2013} reviews empirical work on the usability and
productivity of parallel programming languages and covers much of the earlier
human-subject and programmability literature. Its references include, among
others, the Cowichan studies, early comparisons of parallel programming systems,
Unified Parallel C (UPC) productivity studies, the HPC programmer-productivity experiments of
Hochstein and colleagues, the comparative study of Message Passing Interface (MPI),
UPC, and the X10 programming language by Ebcio\u{g}lu et al., and
subsequent empirical work on concurrent-programming usability.

Wienke et al.~\cite{Wienke2016} provide the second anchor by surveying
software-complexity measures, development-time approaches, software-cost
estimation, and HPC-specific productivity studies. Their bibliography includes
work on UPC productivity, parallel-programming language comparison,
development-time productivity, programmer productivity, HPC productivity
frameworks, performance--effort trade-offs, the Constructive Cost Model (COCOMO), and development-process
instrumentation. 

The review is then extended forward with Harrell et
al.~\cite{Harrell2018}, Pennycook et al.~\cite{Pennycook2021P3}, and Lin et
al.~\cite{Lin2024Productivity}, which represent important attempts to bring
productivity explicitly into the performance-portability discussion. The
forward extension also includes recent language and compiler approaches
designed to improve productive performance portability---JACC,
JACC.Multi, Mojo, JLIR, and Accl++
\cite{ValeroLara2024JACC,ValeroLara2025JACCMulti,Godoy2025Mojo,
Miniskar2026JLIR,Gonzalez2026Accl}---and recent work on AI-assisted HPC
software development~\cite{Teranishi2026AIHPC}. These studies are included
because they expose how the meaning of productivity changes when substantial
portions of portability and optimization are transferred from the programmer
to compiler, runtime, or AI infrastructure.

This selection strategy is appropriate for the objective of this paper because
the goal is not to enumerate every study that has ever used the term
\emph{productivity}. Instead, the objective is to identify the principal
\emph{measurement philosophies} that have influenced HPC productivity research
and to examine how those philosophies evolved.

We classify the methods along the following dimensions:
\begin{enumerate}
    \item \textbf{Measured object:} source-code artifact, development process,
    programmer, or achieved computational outcome.

    \item \textbf{Directness:} direct measurement of effort or productivity
    versus an indirect proxy.

    \item \textbf{Objectivity:} objective, subjective, or mixed.

    \item \textbf{Required data collection:} static source analysis, execution
    measurements, logging, surveys, interviews, or compiler/tool analysis.

    \item \textbf{Reproducibility:} whether another researcher can reproduce
    the score without reproducing the original development process.

    \item \textbf{Ease of use:} computational and experimental cost of obtaining
    the metric.

    \item \textbf{Relationship to performance:} absent, empirically associated,
    or formally incorporated.

    \item \textbf{Relationship to portability:} source portability,
    maintenance burden, platform coverage, or performance portability.

    \item \textbf{Suitability for automated and AI-assisted development:}
    whether the interpretation remains meaningful when substantial amounts of
    code are generated or transformed automatically.
\end{enumerate}

A useful organizing distinction throughout this paper is between
\emph{product measures} and \emph{process measures}. SLOC, code divergence, and
tree-distance metrics measure properties of the produced artifact. Development
time and effort logs measure properties of the process by which the artifact was
obtained. Human-subject measures such as frustration and perceived difficulty
describe properties of the programmer's experience. These quantities may all
contain information relevant to productivity, but they should not be treated as
interchangeable.

\begingroup
\small
\setlength{\tabcolsep}{3pt}
\renewcommand{\arraystretch}{1.12}

\begin{longtable}{@{}
L{0.15\textwidth}
L{0.21\textwidth}
L{0.18\textwidth}
L{0.40\textwidth}
@{}}

\caption{Taxonomy of productivity-related measures reviewed in this
paper. ``Direct'' denotes observation of development effort itself,
whereas ``proxy'' denotes an observable used to infer effort or
productivity. PP denotes performance portability.}
\label{tab:productivity_taxonomy}\\

\toprule
\textbf{Approach} &
\textbf{Measured observable} &
\textbf{Measurement character} &
\textbf{Performance/portability role and main limitation} \\
\midrule
\endfirsthead

\multicolumn{4}{@{}l}{\small\textit{Table~\thetable\ continued}}\\[2pt]
\toprule
\textbf{Approach} &
\textbf{Measured observable} &
\textbf{Measurement character} &
\textbf{Performance/portability role and main limitation} \\
\midrule
\endhead

\midrule
\multicolumn{4}{r@{}}{\small\textit{Continued on next page}}\\
\endfoot

\bottomrule
\endlastfoot


\rowcolor{gray!12}
\multicolumn{4}{@{}l}{\textbf{Static artifact measures}}\\
\addlinespace[2pt]

SLOC/LLOC &
Source-code size &
Proxy; mostly objective &
No explicit performance or portability term. Equal weighting of
semantically unequal code and omission of non-code work make code
quantity a weak representation of effort. \\

\addlinespace[1pt]
\cmidrule(lr){1-4}
\addlinespace[1pt]

Function points &
Functional software size &
Proxy; mixed &
No explicit performance or portability term. Poorly represents
parallelization, optimization, and other HPC-specific work. \\

\addlinespace[1pt]
\cmidrule(lr){1-4}
\addlinespace[1pt]

Halstead complexity &
Operators, operands, and derived complexity &
Proxy; mostly objective &
Provides a richer structural description than SLOC, but the inferred
relationship to human effort has not been validated for HPC. \\

\addlinespace[1pt]
\addlinespace[1pt]

Cyclomatic complexity &
Control-flow complexity &
Proxy; objective &
Can reflect structural complexity but misses optimization, tuning, and
porting activities that do not alter control flow. \\


\addlinespace[4pt]
\rowcolor{gray!12}
\multicolumn{4}{@{}l}{\textbf{Human and process measures}}\\
\addlinespace[2pt]

Human-subject usability &
Time, correctness, difficulty, learning, and debugging &
Partly direct; mixed &
Performance can be included experimentally, but portability is usually
secondary. Results depend strongly on participant experience and study
design. \\

\addlinespace[1pt]
\cmidrule(lr){1-4}
\addlinespace[1pt]

Development time &
Person-hours &
Direct process measure &
Can be related explicitly to performance and may include porting.
Comparison remains strongly dependent on developer experience and
development context. \\

\newpage
\addlinespace[4pt]
\rowcolor{gray!12}
\multicolumn{4}{@{}l}{\textbf{Effort-estimation models}}\\
\addlinespace[2pt]

COCOMO-type models &
Predicted development effort &
Model-based estimate; mixed &
Performance is weak or indirect and portability is not formalized.
Project-size assumptions and calibration are poorly matched to many HPC
development settings. \\

\addlinespace[1pt]
\cmidrule(lr){1-4}
\addlinespace[1pt]

Wienke HPC effort model &
Effort over a performance life cycle &
Model-based estimate; mixed &
Performance is explicit; portability/maintainability appears as an
effort factor. Requires empirical calibration and substantial
development-process data. \\


\addlinespace[4pt]
\rowcolor{gray!12}
\multicolumn{4}{@{}l}{\textbf{Portability-era artifact proxies}}\\
\addlinespace[2pt]

Churn/code changes &
Added, removed, or modified source &
Proxy; objective &
Useful for characterizing development and porting activity, but the
amount of source change need not correspond to human effort. \\

\addlinespace[1pt]
\cmidrule(lr){1-4}
\addlinespace[1pt]

Code divergence/convergence &
Source distance among platform-specific variants &
Proxy; objective &
Jointly analyzed with PP and useful for characterizing specialization
and code sharing. Artifact divergence, however, is not development
effort. \\

\addlinespace[1pt]
\addlinespace[1pt]

Tree-Based Model Divergence (TBMD) &
Tree-based divergence from a selected reference &
Proxy; objective once the baseline and analysis are fixed &
Jointly visualized with PP. Programming models are compared through
their respective divergence from the selected baseline; the result is
baseline-dependent and human effort remains inferred. \\

\end{longtable}

\endgroup


\section{From Usability to Programmer Productivity}
\label{sec:usability}

\subsection{Early Empirical Studies}
\label{subsec:early_empirical}

Much of the early work on productivity in parallel programming was conducted
under the broader heading of \emph{usability}. Parallel programming introduced
difficulties absent or less prominent in sequential development: race
conditions, deadlock, synchronization, communication, load balancing,
nondeterminism, and architecture-specific performance behavior. Consequently,
researchers examined not only the source code produced by programmers but also
the process through which programmers learned, implemented, debugged, and tuned
parallel programs.

The empirical literature reviewed by Marowka~\cite{Marowka2013} illustrates the
diversity of experimental methods. Some studies measured time to obtain the
first correct parallel solution; others separated development time into serial
implementation, parallelization, debugging, and tuning
\cite{Hochstein2005,Ebcioglu2006}. Additional measures included lines of code,
number of characters, parameters, keywords, function calls, correctness, and
subjective ratings. Controlled experiments sometimes instrumented the
development environment, recorded screens, sampled project directories, and
logged compiler invocations and program executions.

These experiments established an important principle that remains relevant
today: productivity cannot be reduced to a single property of the final source
file without potentially losing important information about the development
process. A shorter program may have required greater reasoning, debugging,
learning, or tuning effort than a longer one. Conversely, a verbose low-level
implementation may be generated mechanically or derived from a pre-existing
implementation with relatively little additional human reasoning.

Human-subject studies, however, introduce their own difficulties. Results depend
strongly on participant selection. Many historical studies used students or
programmers inexperienced with the tested programming models, while others used
researchers who had designed the systems being evaluated. Training duration also
varied considerably, from a few hours to an entire semester or longer
\cite{Marowka2013}. Consequently, elapsed time or perceived difficulty can
measure the interaction between a particular person and a programming system
rather than an intrinsic property of the programming system itself.

This distinction is central to the discussion in this paper.
\emph{Programmer productivity} and \emph{programming-model productivity} are
related but not identical concepts. If two programmers with very different
levels of experience require different amounts of time to perform the same
transformation, the measured time cannot be attributed solely to the programming
model.

\subsection{Productivity Cannot Ignore Performance}
\label{subsec:productivity_performance}

An important insight in the early usability literature is that ease of
programming cannot be evaluated independently of the quality of the resulting
program. Marowka~\cite{Marowka2013} explicitly argues that if a language enables
an implementation to be produced easily but the resulting program fails to
scale, then usability alone gives an incomplete assessment. Usability
experiments should therefore include performance objectives.

This observation anticipates later P3 research. A high-level abstraction may
reduce code size and simplify development, yet an implementation that performs
poorly may require additional optimization or even rewriting. Conversely, a
low-level model may require more explicit source code but provide greater
control over optimization. The relevant question is therefore not simply
whether one model requires fewer lines or fewer programming hours than another,
but what useful computational outcome is obtained for the effort invested.

This idea appears more formally in later definitions of productivity as an
output-to-input ratio, discussed later in this review.


\section{Source-Code Size and Static Complexity Measures}
\label{sec:static_metrics}

\subsection{SLOC as the Dominant Proxy}
\label{subsec:sloc}

Source lines of code became the most persistent productivity proxy because of a
compelling practical advantage: they can be counted after development has
finished. No experiment, developer diary, interview, or instrumentation is
required. Given a source repository and a defined counting convention, the
measure is inexpensive and reproducible.

Nguyen et al.~\cite{NguyenSLOC} explain the historical basis of this popularity.
SLOC was a visible product of programming activity and became a major input to
software-cost estimation models including COCOMO, Software Life-Cycle Management
(SLIM), and the System Evaluation and Estimation of Resources--Software Estimation
Model (SEER-SEM). At the
same time, inconsistent definitions of what constitutes a source line motivated
repeated standardization efforts.

The problem is particularly significant in HPC. Wienke et
al.~\cite{Wienke2016} identify several variants in common use: total LOC,
modified LOC, added or removed LOC, uncommented LOC, and counts excluding blank
lines or particular constructs. Different choices may lead to materially
different conclusions. The issue is therefore not simply how many lines a
program contains, but \emph{what is being counted and relative to what
baseline}.

More fundamentally, SLOC assumes a relationship between code quantity and
effort that is often unjustified. Consider two single source lines:
\begin{lstlisting}[language=C]
double x;
\end{lstlisting}
and
\begin{lstlisting}[language=C]
#pragma omp target teams distribute parallel for ...
\end{lstlisting}
Both may contribute one line to a SLOC total, while the second may require the
programmer to reason about data mapping, synchronization, device execution,
correctness, and performance. Wienke et al.~\cite{Wienke2016} use essentially
this example to illustrate why line counts do not reflect the conceptual
complexity of parallel programming.

SLOC also ignores development work that leaves no persistent trace in the final
source file. Changing compiler flags, experimenting with thread or process
placement, profiling, debugging, testing, and evaluating alternative algorithms
can consume substantial effort while leaving the final line count unchanged.
The same is true when existing lines are repeatedly modified while the total
number of lines remains approximately constant. Thus, a stable final SLOC value
does not imply that little development work occurred.

This leads to a first important distinction:
\begin{equation}
\boxed{\text{amount of source code} \neq \text{development effort}}
\label{eq:sloc_not_effort}
\end{equation}
SLOC is an objectively measurable property of an artifact. Development effort
is a property of the process used to produce or transform that artifact.

The distinction becomes even more important in contemporary development
environments. Automated source-to-source migration tools and AI-assisted
programming can generate or transform large volumes of code with comparatively
little direct human editing. Consequently, the historical assumption that the
quantity of generated or modified code provides a reasonable indication of
human programming effort is becoming increasingly difficult to sustain.

The continued use of code size in very recent compiler research illustrates
that SLOC has not disappeared as a productivity proxy. Miniskar et
al.~\cite{Miniskar2026JLIR}, for example, compare the implementation
burden of extending their Julia-native JLIR framework with that of native
C++ MLIR by implementing equivalent dialect functionality and compiler
passes in both systems. They report 67 source lines for the JLIR
implementation and 243 for the C++ MLIR implementation, a factor of
$3.6\times$. This is a useful controlled comparison of implementation
size, but its interpretation remains the same fundamental one discussed
above: fewer source lines describe a smaller software artifact, not
necessarily proportionally less development time, reasoning, debugging,
or learning effort.

\subsection{Function Points}
\label{subsec:function_points}

Function points were developed as an alternative to raw code size by measuring
user-visible functionality through inputs, outputs, inquiries, files, and
interfaces. Wienke et al.~\cite{Wienke2016} note that their independence from a
particular programming language can make them useful early in the development
life cycle. However, this strength becomes a weakness for HPC productivity
analysis: function points do not naturally capture parallelization,
optimization, synchronization, architecture-specific adaptation, or other
non-functional requirements that dominate HPC development effort.

They are therefore better interpreted as a general software-sizing methodology
than as a direct measure of HPC programming productivity.

\subsection{Halstead Complexity}
\label{subsec:halstead}

Halstead's software-science metrics attempt to improve on raw LOC by counting
operators and operands and using them to estimate quantities such as program
volume and programming effort. Such metrics have occasionally been used to
compare parallel programming approaches. However, Wienke et
al.~\cite{Wienke2016} identify two important difficulties. Definitions of
operators and operands can vary among programming models, particularly for
directive-based approaches, and the conversion from source-level complexity to
programming time relies on assumptions about human cognitive processing that
have not been validated for HPC development.

Halstead metrics are therefore more structurally informed than SLOC, but they
continue to infer development effort from characteristics of the finished source
artifact.

\subsection{Cyclomatic Complexity}
\label{subsec:cyclomatic}

McCabe's cyclomatic complexity measures the number of independent paths through
a program's control-flow graph~\cite{McCabe1976}. Vanderwiel et
al.~\cite{Vanderwiel1998} used the measure to compare the relative complexity
and performance of parallel programming approaches. However, as Wienke et
al.~\cite{Wienke2016} observe, many important HPC optimization activities do
not necessarily alter control flow. Compiler options, affinity decisions,
thread binding, and other tuning operations may require substantial effort while
leaving cyclomatic complexity unchanged.

Again, the fundamental limitation is not that cyclomatic complexity is
incorrectly defined. Rather, software structural complexity and programmer
effort are different quantities.


%
%
%
%


\section{Development Time and Effort Models}
\label{sec:effort_models}

\subsection{Development Time as a Process Measure}
\label{subsec:development_time}

The limitations of source-code metrics motivated a shift from properties of the
software artifact toward properties of the development process itself.
Development time, usually expressed in person-hours or person-months, is one of
the most direct quantities available for this purpose. Unlike SLOC or static
complexity, development time can in principle include work that leaves no
persistent trace in the final source code, including learning a programming
model, understanding an existing application, debugging, testing, profiling,
performance tuning, experimenting with compiler options, and porting to a new
architecture.

Wienke et al.~\cite{Wienke2016} therefore identify development time as a
particularly appropriate quantitative effort measure for HPC. They argue that
the activities relevant to parallel development should include familiarization
with the programming model, hardware, existing domain code, and development
tools, as well as parallelization, porting, tuning, correctness testing,
debugging, and preparation of the execution environment. This broader
interpretation addresses one of the central weaknesses of code-size measures:
substantial development activity can occur without increasing the size of the
program.

Development time, however, introduces a different set of difficulties. It is
not always clear which activities should be included, and elapsed development
time is strongly affected by programmer capability, prior knowledge,
familiarity with the programming model, the application, and the target
architecture. The same programming task can therefore require substantially
different effort when performed by a novice and by an experienced developer.
Development time is consequently closer to the human effort of interest than
SLOC, but it is not an intrinsic property of a programming model.

The practical collection of development-time data is itself difficult. Manual
developer diaries require sustained participation and are vulnerable to missing
or inconsistent entries. Automatic instrumentation can reduce this burden, but
typically observes only activities performed through instrumented tools and may
miss time spent thinking, reading, planning, or discussing a problem
\cite{Wienke2016}. Thus, measuring effort more directly improves construct
validity at the cost of substantially greater experimental complexity.

\subsection{Relative Development Time Productivity}
\label{subsec:rdtp}

A more explicit attempt to connect development effort with computational outcome
is the Relative Development Time Productivity (RDTP) measure introduced by
Funk et al.~\cite{Funk2005}. In the notation commonly used to describe the
measure,
\begin{equation}
    \mathrm{RDTP}
    =
    \frac{\mathrm{speedup}}
         {\mathrm{relative\ development\ effort}} .
\label{eq:rdtp}
\end{equation}
This formulation is important because it does not equate productivity with
effort alone. Instead, productivity is interpreted as an output-to-input ratio:
the computational improvement obtained for the development effort invested.
Harrell et al.~\cite{Harrell2018} later adopted the same general interpretation,
writing productivity as
\begin{equation}
    \mathrm{productivity}
    =
    \frac{\mathrm{output}}
         {\mathrm{input}},
\label{eq:productivity_general}
\end{equation}
where, for an individual programmer, application performance can represent the
output and development effort the input.

The appeal of Eq.~\eqref{eq:rdtp} is clear: more development effort is not
automatically interpreted as higher productivity, nor is high performance
sufficient by itself. Nevertheless, several problems remain. First, both the
performance output and the effort denominator require a clearly defined
baseline. Second, development effort remains dependent on the developer and
experimental setting. Third, the relationship between development effort and
performance is generally nonlinear. Harrell et al.~\cite{Harrell2018}
explicitly caution that productivity measured at a particular operating point
should not be extrapolated as if additional effort would produce proportional
performance improvement.

Thus, RDTP provides an important conceptual step by directly relating effort to
an achieved computational outcome, but it does not eliminate the difficulties
of defining and measuring the effort term.

\subsection{From COCOMO to an HPC Performance Life Cycle}
\label{subsec:wienke_model}

Wienke et al.~\cite{Wienke2016} develop a more comprehensive effort-estimation
methodology by adapting concepts from software-cost estimation, particularly
COCOMO II, to HPC. Conventional COCOMO-type models primarily relate software
size and a collection of cost drivers to predicted development effort.
Wienke et al. argue that this formulation is poorly matched to HPC, where
developers often begin with an existing scientific code and where considerable
effort may be spent obtaining the final increments of performance without large
changes in source-code size.

Their central modification is to replace project size as the principal base
quantity with a \emph{performance life cycle}. The proposed model has the form
\begin{equation}
    E = S\, f(P)^{R},
\label{eq:wienke_effort}
\end{equation}
\mbox{where $E$ denotes} development \mbox{effort, $P$ represents} achieved performance,
$f(P)$ describes the relationship between performance and effort, and $S$ and
$R$ incorporate additional effort drivers. The model can represent familiar
optimization behavior such as an 80--20 relationship, but the authors emphasize
that more complicated performance life cycles may occur in practice.

This formulation is significant for the history of HPC productivity research
because performance is no longer merely reported beside development effort.
Instead, achieved performance becomes an explicit component of the effort model.
The methodology also recognizes that effort is influenced by multiple factors.
Based on their initial data, Wienke et al. identify developer pre-knowledge,
the type of code work, the programming model together with compiler support,
and the required performance level among the most important factors
\cite{Wienke2016}.

Portability also appears in their analysis, although not in the later formal
sense of performance portability. Portability and maintainability over the
lifetime of a code are included as one possible development-effort factor,
representing the need for software to remain usable across changing
architectures. In their initial factor analysis, however, portability and
maintainability were ranked below factors such as pre-knowledge, code-work type,
programming-model/compiler support, and performance. The methodology therefore
recognizes portability as a potential contributor to development effort but does
not define a mathematical relationship between productivity and a
performance-portability metric.

\subsection{EffortLog and the Practical Measurement Problem}
\label{subsec:effortlog}

To obtain the data required by the performance-life-cycle model, Wienke et al.
developed EffortLog, an electronic developer diary that periodically records
how development time is spent and associates that effort with performance
milestones~\cite{Wienke2016}. Unlike source-code instrumentation alone,
EffortLog can include categories such as thinking, testing, debugging, and
activities performed away from the editor. It can also record information on
compiler flags, hardware, programming models, and achieved performance.

EffortLog illustrates both the strength and the difficulty of direct effort
measurement. It captures substantially more of the development process than a
static code metric, but still depends on developer participation and correct
classification of activities. Logging introduces overhead, developers may
forget or postpone entries, and the meaning of a reported hour remains dependent
on experience and context.

The Wienke methodology should therefore be distinguished from a simple
productivity metric. It has a formal mathematical structure and a statistical
calibration methodology, but it is not a universal closed-form measure that can
be computed from an arbitrary HPC codebase. The function $f(P)$ and the effort
drivers must be established empirically, and meaningful estimation requires
sufficient development-process data. This makes the approach particularly
valuable for understanding and predicting effort in real development projects,
but considerably more demanding than the lightweight metrics normally used in
benchmark-oriented performance-portability studies.


\section{From Productivity to P3}
\label{sec:harrell}

\subsection{Performance Portability in Brief}
\label{subsec:pp_background}

Before discussing attempts to incorporate productivity into performance
portability, it is useful to summarize the underlying concept. Performance
portability concerns the ability of an implementation to achieve useful
performance across a specified set of target platforms. Quantitative
performance-portability metrics therefore commonly begin with a
platform-specific performance efficiency and aggregate those efficiencies
over the platform set.

Two forms of efficiency are frequently used. \emph{Architectural efficiency}
measures achieved performance relative to a hardware-oriented reference,
such as theoretical peak performance or an appropriate architectural
performance bound. \emph{Application efficiency}, by contrast, measures the
performance of the implementation of interest relative to a reference
implementation on the same platform, for example a highly optimized native
implementation or vendor library. Thus, architectural efficiency asks how
effectively the hardware is being utilized, whereas application efficiency
asks how closely a portable implementation approaches an appropriate
platform-specific performance reference. These alternatives and their
implications for performance-portability metrics are discussed in more
detail in~\cite{Marowka2023PPComparison}.

Let $H$ denote the target platform set and let
$\eta(a,p,h)$ denote the efficiency of implementation $a$ solving problem
$p$ on platform $h$. As one example, the
\funnyPbar{} metric aggregates the observed efficiencies
using the arithmetic mean over the supported subset $S\subseteq H$:

\begin{equation}
\funnyPbar(a,p,S,H)
=
\begin{cases}
\dfrac{1}{|S|}
\displaystyle\sum_{h\in S}\eta(a,p,h),
& |S|>0,\\[2ex]
0, & |S|=0.
\end{cases}
\label{eq:ppbar_background}
\end{equation}

The resulting value therefore summarizes the average efficiency achieved
on the supported platforms. The size and composition of $S$ remain important
context: an aggregate efficiency score should not be interpreted independently
of the platform coverage on which it was obtained.

Other formulations aggregate the same underlying efficiency information
differently. In particular, the original \funnyP{} metric introduced by Pennycook
et al. uses a harmonic-mean formulation and assigns a score of zero when the
implementation is unsupported on any platform in the specified set, whereas
\funnyPbar{} uses the arithmetic mean over the supported
subset. A detailed mathematical and empirical comparison of \funnyP{} and
\funnyPbar{} is provided by Marowka
\cite{Marowka2023PPComparison}; those differences are not repeated here
because the present review is concerned primarily with how productivity has
been incorporated into performance-portability analysis.

For the discussion that follows, the important point is that a
performance-portability metric is derived from observed performance
efficiencies across platforms. Productivity measures such as development
time, code divergence, code convergence, and TBMD are derived from different
observables. The central question in the following sections is therefore how,
and whether, these productivity-related quantities can be meaningfully
related to performance portability.

\subsection{Adding Productivity to Performance Portability}
\label{subsec:harrell_p3}

Harrell et al.~\cite{Harrell2018} mark an important transition from studying
productivity largely as a separate software-development concern to considering
it explicitly together with performance and portability. Their motivation is
that a performance-portable implementation may achieve high efficiency across
multiple platforms while requiring very different amounts of development and
maintenance effort. In particular, a high performance-portability score can in
principle be obtained using separate, highly optimized implementations for each
platform, even though maintaining those implementations may be expensive.

The study therefore distinguishes two broad approaches to measuring developer
effort. Direct methods record the development process itself, for example
through developer diaries or effort logs. Indirect methods estimate effort from
observable properties of the resulting code, including SLOC and the number of
code changes. This distinction is particularly useful because it makes explicit
a recurring issue in the productivity literature: quantities that are easy to
measure after development are often proxies rather than direct observations of
effort.

Harrell et al. extend simple code-change counts by introducing the concept of
\emph{code divergence}. Let $A$ denote a set of implementations or ports of an
application and let $d(a_i,a_j)$ denote a distance between two versions. Their
divergence measure is the average pairwise distance,
\begin{equation}
    D(A)
    =
    { {|A| \choose 2} }^{-1}
    \sum_{\{a_i,a_j\}\subset A}
    d(a_i,a_j).
\label{eq:harrell_divergence}
\end{equation}
A low divergence corresponds to implementations that remain structurally close
to one another, while greater divergence indicates increasingly different
platform-specific versions. This provides a useful measure of one contributor
to maintenance burden: when several distinct representations of the same
functionality must be maintained, a future modification may need to be repeated
across multiple code paths.

The paper goes further and conceptually relates source distance to development
cost through transformation-dependent functions. This is an important
recognition that code distance and human effort are not identical: the same
amount of source-code change may have very different costs depending on the
transformation, tools, and programming model involved.

The accompanying case studies collect several forms of productivity-related
information, including development hours, net line changes, churn, divergence,
and subjective difficulty or frustration, together with performance and
performance-portability measurements. The Truchas study, for example, records
development time, net line changes, and cumulative frustration separately for
OpenMP central processing unit (CPU), OpenMP graphics processing unit (GPU)
offload, and CUDA implementations. Such data are valuable
because they expose dimensions of the development process that are invisible in
the final performance result.

At the same time, the study demonstrates the difficulty of constructing a
single quantitative productivity measure from these observations. The generic
output/input formulation and RDTP are discussed, but the case studies primarily
report the effort-related variables and performance-portability results
separately. Harrell et al. themselves characterize the productivity results as
mixed and emphasize the scarcity of reliable effort data across developers,
applications, and programming approaches~\cite{Harrell2018}.

The contribution of this work is therefore best understood as a methodology for
collecting and comparing quantities relevant to productive performance
portability rather than as a completed mathematical integration of the three
Ps. Code divergence, development time, code churn, and subjective difficulty
describe different aspects of development. They can help explain why two
performance-portable solutions differ in maintenance or development cost, but
none of them is mathematically determined by the performance-portability score.


\section{Code Divergence, Code Convergence, and Performance Portability}
\label{sec:pennycook_p3}

Pennycook et al.~\cite{Pennycook2021P3} make a more explicit attempt to bring
productivity into the same quantitative analysis as performance portability.
Their starting point is the observation that a performance-portability metric
indicates whether an implementation delivers efficient execution across a
platform set, but does not explain \emph{how} that result was obtained. Two
implementations may therefore achieve similar performance-portability scores
while relying on very different degrees of platform-specific specialization.

The authors also acknowledge a fundamental difficulty: measuring programmer
productivity in general and absolute terms is unlikely to be practical.
Instead, they focus on quantities that are known or assumed to influence
productivity. In particular, platform-specific development effort is
approximated by the amount of platform-specific code that an application
contains. The underlying reasoning is that specialization creates two potential
costs: new code may be required when support for another platform is added, and
multiple existing representations of the same functionality must subsequently
be maintained.

\subsection{Code Divergence and Code Convergence}
\label{subsec:code_convergence}

For application $a$, problem $p$, and platform set $H$, Pennycook et al. define
Code Divergence (CD) as an average pairwise source-code distance,
\begin{equation}
    CD(a,p,H)
    =
    { {|H| \choose 2} }^{-1}
    \sum_{\{i,j\}\subset H}
    d_{i,j}(a,p),
\label{eq:code_divergence}
\end{equation}
where $d_{i,j}$ measures the distance between the source code required for
platforms $i$ and $j$. Their implementation uses a Jaccard distance over the
sets of source lines required by the two platforms,
\begin{equation}
    d_{i,j}(a,p)
    =
    1 -
    \frac{|c_i(a,p)\cap c_j(a,p)|}
         {|c_i(a,p)\cup c_j(a,p)|}.
\label{eq:jaccard_code}
\end{equation}
The corresponding Code Convergence (CC) quantity is
\begin{equation}
    CC(a,p,H)=1-CD(a,p,H),
\label{eq:code_convergence}
\end{equation}
so that larger values of both performance portability and code convergence are
interpreted as desirable.

This construction has several practical advantages. It is objective once the
source-selection and preprocessing rules are fixed, it can be automated, and it
provides information about platform specialization and potential maintenance
burden that is not present in a performance-portability score alone.

\subsection{The Performance-Portability--Code-Convergence Plane}
\label{subsec:pcc_plane}

Pennycook et al. plot performance portability and code convergence in a
two-dimensional PP--CC navigation chart. The upper-right region represents the
desired combination of high performance portability and high source-code
convergence. Movements through this plane can then be interpreted in software
engineering terms. Platform-specific optimization may improve performance
portability while decreasing convergence, whereas introducing an abstraction
may increase convergence while preserving or, in some cases, reducing
performance.

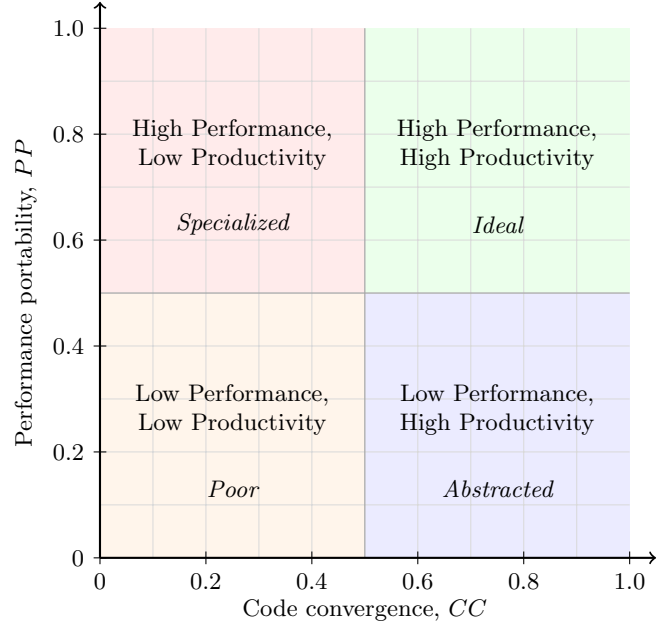
\begin{figure}[t]
\centering
\begin{tikzpicture}[x=7cm,y=7cm]

\fill[red!8]    (0,0.5) rectangle (0.5,1);   
\fill[green!8]  (0.5,0.5) rectangle (1,1);   
\fill[orange!8] (0,0) rectangle (0.5,0.5);   
\fill[blue!8]   (0.5,0) rectangle (1,0.5);   

\draw[step=0.1, very thin, gray!30] (0,0) grid (1,1);

\draw[thin, gray!70] (0.5,0) -- (0.5,1);
\draw[thin, gray!70] (0,0.5) -- (1,0.5);

\draw[->,thick] (0,0) -- (1.05,0);
\draw[->,thick] (0,0) -- (0,1.05);

\foreach \x in {0,0.2,0.4,0.6,0.8,1.0}
{
    \draw (\x,0.012) -- (\x,-0.012);
    \node[below,font=\scriptsize] at (\x,-0.012) {\x};
}

\foreach \y in {0,0.2,0.4,0.6,0.8,1.0}
{
    \draw (0.012,\y) -- (-0.012,\y);
    \node[left,font=\scriptsize] at (-0.012,\y) {\y};
}

\node[below=12pt] at (0.5,0) {\small Code convergence, $CC$};
\node[rotate=90] at (-0.14,0.5) {\small Performance portability, $PP$};

\node[align=center,font=\small] at (0.25,0.78)
    {High Performance,\\Low Productivity};
\node[align=center,font=\scriptsize\itshape] at (0.25,0.63)
    {Specialized};

\node[align=center,font=\small] at (0.75,0.78)
    {High Performance,\\High Productivity};
\node[align=center,font=\scriptsize\itshape] at (0.75,0.63)
    {Ideal};

\node[align=center,font=\small] at (0.25,0.28)
    {Low Performance,\\Low Productivity};
\node[align=center,font=\scriptsize\itshape] at (0.25,0.13)
    {Poor};

\node[align=center,font=\small] at (0.75,0.28)
    {Low Performance,\\High Productivity};
\node[align=center,font=\scriptsize\itshape] at (0.75,0.13)
    {Abstracted};

\end{tikzpicture}

\caption{Schematic performance-portability--code-convergence
(PP--CC) plane. Higher code convergence is used here as a proxy for
greater productivity. The four quadrants summarize the qualitative
trade-off between performance portability and productivity. The diagram
is conceptual and does not represent experimental data.}
\label{fig:pcc_plane}
\end{figure}

The quadrant names in Fig.~\ref{fig:pcc_plane} are intended only as
qualitative shorthand. \emph{Ideal} denotes implementations that combine
high performance portability with high code convergence. \emph{Specialized}
denotes implementations that achieve high performance portability at the
cost of lower convergence, typically reflecting greater platform-specific
specialization. \emph{Abstracted} denotes implementations with relatively
high convergence but limited performance portability, suggesting a
maintainable common source that has not yet translated into sufficiently
portable performance. Finally, \emph{Poor} denotes implementations that
are low in both performance portability and convergence.

As illustrated in Fig.~\ref{fig:pcc_plane}, this construction is useful
as a decision-support visualization. It makes two
desirable but distinct properties visible simultaneously and highlights the
consequences of specialization and abstraction. It also avoids collapsing the
two quantities immediately into one scalar.

However, plotting two quantities on orthogonal axes does not by itself establish
a mathematical relationship between them. Performance portability is derived
from observed performance efficiencies across a platform set, whereas code
convergence is derived from differences among the source representations used
for those platforms. One quantity does not determine the other. Two
implementations can have identical source convergence and very different
performance, or similar performance-portability scores and very different
source structures.

This distinction becomes particularly important for single-source programming
models. An OpenMP, Kokkos, SYCL, or other implementation may use one common
source representation across the complete platform set while its achieved
performance varies substantially among platforms. In such a case, source
convergence can remain approximately constant while performance portability
changes through compiler, runtime, algorithmic, or hardware effects.

\subsection{Combining the Quantities}
\label{subsec:pcc_scalar}

Pennycook et al. also consider a scalar aggregation
\begin{equation}
    M(a,p,H)
    =
    \min\left(
        \mathcal{P}(a,p,H),
        CC(a,p,H)
    \right),
\label{eq:pcc_min}
\end{equation}
where $\mathcal{P}$ denotes the chosen performance-portability metric.

This construction is easy to interpret: a high value requires both high
performance portability and high convergence. Nevertheless, Eq.~\eqref{eq:pcc_min}
should be interpreted as an aggregation rule rather than as evidence of a
derived relationship between the two underlying quantities. Their common
normalization to the interval $[0,1]$ and the fact that ``higher is better'' do
not make performance efficiency and source-code convergence physically or
mathematically commensurate.

Consequently, the PP--CC framework represents an important advance in joint P3
analysis, especially as a visualization and development-navigation tool, but it
does not resolve the more fundamental question of how programmer productivity
is formally related to performance portability.


\section{Semantic Divergence and Tree-Based Model Divergence}
\label{sec:tbmd}

Lin, Deakin, and McIntosh-Smith~\cite{Lin2024Productivity} address an important
weakness of SLOC and textual code divergence: two programs can differ greatly
in formatting or surface syntax while expressing similar semantics, whereas a
small textual modification can introduce an important semantic change. Their
Tree-Based Model Divergence (TBMD) approach therefore moves the comparison from
raw source text toward compiler-derived representations of the program.

The authors explicitly restrict their interpretation of productivity to the
\emph{codebase artifact}. They do not attempt to measure programmer behavior,
experience, or perceived difficulty, which they recognize as potentially
subjective~\cite{Lin2024Productivity}. This restriction gives TBMD a clearer
measurement target than approaches that mix human and artifact properties, but
it also means that TBMD should not be interpreted as a direct measurement of
human development effort.

\subsection{From Textual Difference to Semantic Difference}
\label{subsec:tbmd_trees}

The method represents each codebase at several levels.
$T_{\mathrm{src}}$ is a normalized syntax-oriented representation derived
from the concrete syntax tree, $T_{\mathrm{sem}}$ captures frontend
semantic information from the abstract syntax tree, and
$T_{\mathrm{ir}}$ represents lower-level, platform-independent
intermediate-representation semantics. These quantities are therefore
\emph{tree representations}, rather than productivity scores by themselves.
For each representation, Tree Edit Distance (TED) is used to quantify the
transformation required between a selected reference codebase and a
programming-model implementation.

In the principal productivity formulation of Lin et al., the reference is
a serial codebase $S$. Each parallel implementation $P_M$, produced using
programming model $M$, is compared separately with $S$. Thus,
$T_{\mathrm{src}}$, $T_{\mathrm{sem}}$, and $T_{\mathrm{ir}}$ give
different views of the divergence of $P_M$ from the same reference.
Comparisons among programming models are consequently indirect: for
example, Kokkos and OpenMP are compared by comparing their respective
divergences from $S$, rather than by directly measuring a Kokkos--OpenMP
distance.
Let $B$ denote the selected baseline codebase and let $P_M$ denote the
implementation produced using programming model $M$. Corresponding source
units in $B$ and $P_M$ are first matched. A source unit consists of a
source file together with its relevant dependencies. Let
$\mathcal{M}(B,P_M)=\{(F_i,G_i)\}_{i=1}^{m}$ denote the set of $m$
matched source-unit pairs, where $F_i$ belongs to the baseline and $G_i$
to the programming-model implementation. For a selected tree
representation $T$, the aggregate tree distance is

\begin{equation}
    d_T(B,P_M)
    =
    \sum_{i=1}^{m}
    d_{\mathrm{TED}}\!\bigl(T(F_i),T(G_i)\bigr).
\label{eq:tbmd_distance}
\end{equation}

Here, $T$ may denote $T_{\mathrm{src}}$, $T_{\mathrm{sem}}$, or
$T_{\mathrm{ir}}$, thereby producing source-oriented or semantic views of
the transformation from the baseline to $P_M$. The resulting distances
are normalized to obtain comparable model-divergence values.

Consequently, two programming models $M_1$ and $M_2$ are normally
compared through $D_T(B,P_{M_1})$ versus $D_T(B,P_{M_2})$, rather than
through a direct distance between $P_{M_1}$ and $P_{M_2}$.

This is a substantial improvement over SLOC when the objective is to
characterize the structure of a transformation. Rather than treating every
source line as an equivalent unit, TBMD can distinguish changes at different
levels of representation and expose cases where perceived source complexity and
compiler-visible semantic complexity differ.

\subsection{The Serial Baseline and the Meaning of Productivity}
\label{subsec:tbmd_baseline}

The interpretation of TBMD is inherently relative to the selected
reference implementation. In the principal formulation of Lin et al.,
the serial codebase $S$ serves as that reference and is treated as
representing maximum productivity. Each parallel implementation $P_M$ is
then evaluated according to its divergence from $S$, rather than through
direct pairwise comparison with the other programming models.

The farther $P_M$ lies from the serial reference according to the chosen
tree representation, the greater the inferred transformation and,
according to the productivity interpretation of the metric, the greater
the presumed porting effort.

This construction provides a reproducible relative measure, but its
interpretation as productivity embeds an important assumption:
\begin{equation}
    \text{greater baseline divergence}
    \Longrightarrow
    \text{greater presumed effort}.
\label{eq:tbmd_assumption}
\end{equation}
That assumption is more sophisticated than equating effort with SLOC, because
the transformation is characterized semantically rather than merely textually.
Nevertheless, it remains an inference from a property of the resulting artifact
to the effort required to produce it.

The choice of baseline is also consequential. Many mature HPC applications are
not developed by starting from a clean serial implementation; they may be
parallel from their inception or may evolve from an existing CUDA, MPI,
OpenMP, or other implementation. Lin et al. themselves demonstrate that
divergence changes when the starting point is changed from serial code to CUDA
code. TBMD therefore measures transformation relative to a selected baseline,
not an absolute productivity property of a programming model.

\subsection{TBMD and Performance Portability}
\label{subsec:tbmd_pp}

Lin et al. combine TBMD with performance portability by plotting the two
quantities in an enhanced navigation chart inspired by the earlier PP--CC
framework. Results are presented for programming-model variants of CloverLeaf
and TeaLeaf, with source-oriented and semantic divergence shown together with
performance portability~\cite{Lin2024Productivity}.

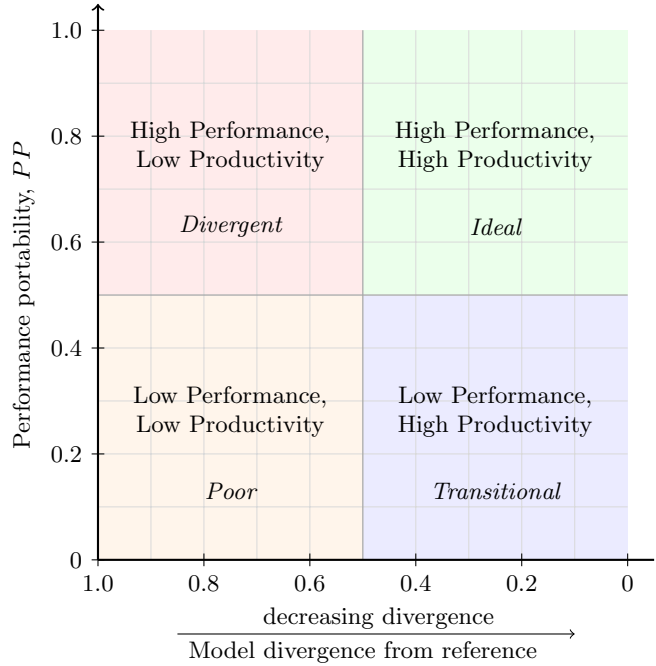
\begin{figure}[t]
\centering
\begin{tikzpicture}[x=7cm,y=7cm]

\fill[red!8]    (0,0.5) rectangle (0.5,1);   
\fill[green!8]  (0.5,0.5) rectangle (1,1);   
\fill[orange!8] (0,0) rectangle (0.5,0.5);   
\fill[blue!8]   (0.5,0) rectangle (1,0.5);   

\draw[step=0.1, very thin, gray!30] (0,0) grid (1,1);

\draw[thin, gray!70] (0.5,0) -- (0.5,1);
\draw[thin, gray!70] (0,0.5) -- (1,0.5);

\draw[thick] (0,0) -- (1.05,0);
\draw[->,thick] (0,0) -- (0,1.05);

\foreach \x/\lbl in {0/1.0,0.2/0.8,0.4/0.6,0.6/0.4,0.8/0.2,1.0/0}
{
    \draw (\x,0.012) -- (\x,-0.012);
    \node[below,font=\scriptsize] at (\x,-0.012) {\lbl};
}

\foreach \y in {0,0.2,0.4,0.6,0.8,1.0}
{
    \draw (0.012,\y) -- (-0.012,\y);
    \node[left,font=\scriptsize] at (-0.012,\y) {\y};
}

\draw[->,thin] (0.15,-0.14) -- (0.90,-0.14);
\node[font=\scriptsize] at (0.53,-0.11) {decreasing divergence};

\node[below=27pt,align=center] at (0.5,0)
    {\small Model divergence from reference};
\node[rotate=90] at (-0.14,0.5)
    {\small Performance portability, $PP$};

\node[align=center,font=\small] at (0.25,0.78)
    {High Performance,\\Low Productivity};
\node[align=center,font=\scriptsize\itshape] at (0.25,0.63)
    {Divergent};

\node[align=center,font=\small] at (0.75,0.78)
    {High Performance,\\High Productivity};
\node[align=center,font=\scriptsize\itshape] at (0.75,0.63)
    {Ideal};

\node[align=center,font=\small] at (0.25,0.28)
    {Low Performance,\\Low Productivity};
\node[align=center,font=\scriptsize\itshape] at (0.25,0.13)
    {Poor};

\node[align=center,font=\small] at (0.75,0.28)
    {Low Performance,\\High Productivity};
\node[align=center,font=\scriptsize\itshape] at (0.75,0.13)
    {Transitional};

\end{tikzpicture}

\caption{Schematic performance-portability--TBMD plane. The
horizontal axis is shown in decreasing-divergence order, so movement to
the right indicates greater similarity to the reference codebase and is
used here as a proxy for greater productivity. The four quadrants
summarize the qualitative trade-off between performance portability and
productivity. The diagram is conceptual and does not represent
experimental data.}
\label{fig:pp_tbmd_plane}
\end{figure}

The quadrant names in Fig.~\ref{fig:pp_tbmd_plane} again serve only as
qualitative shorthand. \emph{Ideal} denotes implementations that combine
high performance portability with low model divergence from the chosen
reference codebase. \emph{Divergent} denotes implementations that attain
high performance portability but only through substantial divergence from
that reference. \emph{Transitional} denotes implementations that remain
relatively close to the reference codebase but have not yet achieved high
performance portability, suggesting an intermediate stage in the porting
or optimization process. Finally, \emph{Poor} denotes implementations
that are low in performance portability and also highly divergent from
the reference.

As summarized schematically in Fig.~\ref{fig:pp_tbmd_plane}, this joint
representation provides richer information than a SLOC-based comparison.
In particular, the distinction between perceived source complexity
and semantic complexity can reveal cases in which a programming model expresses
parallelism concisely at the source level while still introducing significant
semantic structure.

However, the underlying conceptual issue remains similar to that of the PP--CC
framework. Performance portability and TBMD are computed from different
observables. TBMD measures transformation of the codebase relative to a
reference; performance portability measures execution efficiency over a
platform set. Displaying them in a common plane is informative, but it does not
establish that one is mathematically derived from the other.

There are also practical limitations. The method requires compiler-derived tree
representations and a nontrivial analysis infrastructure, and Tree Edit Distance
can become computationally and memory intensive for large applications.
Furthermore, the initial implementation assigns equal cost to tree-edit
operations, even though different semantic transformations may require very
different amounts of human reasoning. The method therefore substantially
improves the characterization of source transformation while still leaving the
mapping from transformation magnitude to human productivity unresolved.


\section{Contemporary Language and Compiler Approaches to Productive Performance Portability}
\label{sec:productive_compilers}

\subsection{Multi-Level Abstraction and Mojo}
\label{subsec:mojo_mlir}

Recent compiler infrastructure changes the productivity discussion by
moving part of the architecture-specific work from application programmers
into reusable compiler representations and transformations. MLIR is a
prominent example. Rather than forcing all compilation stages into one
fixed intermediate representation, MLIR provides extensible dialects and
multiple levels of representation through which high-level program
semantics can be progressively lowered toward target-specific code
\cite{Lattner2021MLIR}. From a productivity perspective, this architecture
is important because transformations that would otherwise be implemented
separately in application code or individual compiler back ends can instead
be represented within reusable compiler infrastructure. MLIR itself,
however, is compiler infrastructure rather than a productivity metric.

Godoy et al.~\cite{Godoy2025Mojo} examine Mojo, a language built on
MLIR that explicitly aims to narrow both performance and productivity
gaps while retaining interoperability with the Python ecosystem. Their
study evaluates scientific GPU kernels on NVIDIA and AMD devices,
compares performance with CUDA and HIP baselines, and summarizes the
results using a performance-portability metric. Productivity, by contrast,
is motivated primarily through reuse of the Python ecosystem,
interoperability, and the possibility of expressing vendor-independent GPU
code within one language environment.

This asymmetry is informative for productivity measurement. Performance
and performance portability are evaluated quantitatively, whereas the
productivity benefit is inferred largely from properties of the language
and software ecosystem. The authors themselves note that Mojo's GPU
programming model remains relatively low level and that its learning curve
is nontrivial. Thus, a language may plausibly reduce ecosystem
fragmentation while the amount of human effort saved remains difficult to
express as a general quantitative productivity value.

\subsection{JACC, JACC.Multi, and JLIR}
\label{subsec:jacc_jlir}

Valero-Lara et al.~\cite{ValeroLara2024JACC} pursue a similar objective
through JACC (Julia for ACCelerators), which provides a lightweight
high-level Julia front end over multiple CPU and GPU back ends. The same
Julia source can therefore be mapped to heterogeneous targets without
requiring the application programmer to maintain separate vendor-specific
implementations. Their evaluation emphasizes performance relative to
native or vendor-oriented solutions while describing the common front end
as reducing the programmability cost of accelerator support.

JACC.Multi extends this idea to multi-GPU execution
\cite{ValeroLara2025JACCMulti}. GPU-to-GPU communication is incorporated
into the programming framework so that applications can target NVIDIA and
AMD multi-GPU systems without exposing all communication details at the
application level. Here again, productivity is principally an engineering
objective: reducing the amount of target-specific programming required to
obtain a portable high-performance implementation.

JLIR moves the abstraction boundary further into the compiler
\cite{Miniskar2026JLIR}. It provides a Julia-native, MLIR-inspired
intermediate representation and can recognize parallel and reduction
patterns in ordinary serial Julia code and lower them automatically to
JACC kernels without requiring user annotations. Consequently, a large
difference between the programmer-visible source and the executed
accelerator implementation may be created by compiler transformation
rather than by corresponding manual source modification.

JLIR is particularly relevant to the history reviewed in this paper
because it also contains an explicit source-size comparison, discussed in
Section~\ref{subsec:sloc}. The return to SLOC in a modern compiler
framework illustrates the continuing practical attraction of artifact
measures: they are easy to obtain and reproduce even when the underlying
concept of human productivity remains more difficult to observe directly.

\subsection{Accl++ and Runtime Compilation}
\label{subsec:accl}

Gonzalez et al.~\cite{Gonzalez2026Accl} approach the same problem through
Accl++, an architecture-agnostic programming interface embedded in C++
and supported by runtime compilation. The same application binary can
target heterogeneous devices from different vendors, while device-specific
code generation is deferred to the runtime compilation infrastructure.
Their evaluation reports relatively small runtime-compilation overheads
and performance close to native CUDA, HIP, or OpenCL implementations on
the studied systems.

For productivity analysis, the important feature is not merely runtime
compilation itself but the redistribution of development work. Code
portability can reduce the need to rewrite or maintain separate
vendor-specific implementations, yet the study does not directly measure
the human effort saved by doing so. As with JACC and Mojo, improved
programmability and reduced platform-specific coding provide evidence of
a productivity-oriented design, but they are not themselves direct
measurements of programmer productivity.

\subsection{Engineering for Productivity Versus Measuring Productivity}
\label{subsec:engineering_vs_measurement}

Table~\ref{tab:modern_productivity_systems} summarizes how these recent
systems relate to productivity measurement.

\begin{table}[t]
\caption{Recent language and compiler systems designed to improve productive
performance portability. The final column distinguishes the mechanism used
to improve productivity from a direct productivity measurement.}
\label{tab:modern_productivity_systems}
\small
\begin{tabularx}{\textwidth}{@{}p{0.13\textwidth}YYY@{}}
\toprule
\textbf{System} &
\textbf{Main mechanism} &
\textbf{Performance/portability evaluation} &
\textbf{Productivity treatment} \\
\midrule

JACC / JACC.Multi &
Common high-level Julia front end and heterogeneous back ends; multi-GPU
abstraction &
Performance across CPU/GPU and NVIDIA/AMD multi-GPU systems &
Productivity inferred from programmability, common source, and reduced
target-specific implementation burden; no direct human-effort metric. \\

\addlinespace

Mojo &
MLIR-based compilation and Python interoperability &
Vendor baselines on NVIDIA and AMD GPUs; performance portability evaluated
quantitatively &
Productivity associated with ecosystem reuse and language integration;
learning and low-level programming requirements remain relevant. \\

\addlinespace

JLIR &
Julia-native MLIR-inspired IR and automatic JACC kernel extraction &
Generated CPU/GPU kernels compared with serial and accelerator baselines &
Automatic transformation reduces explicit programming; one productivity
comparison uses SLOC, remaining an artifact proxy. \\

\addlinespace

Accl++ &
Architecture-agnostic C++ interface and runtime compilation &
Same application across heterogeneous vendor devices; native-oriented
performance comparison &
Productivity inferred from code portability and reduced vendor-specific
rewriting rather than directly measured development effort. \\

\bottomrule
\end{tabularx}
\end{table}

These systems constitute an important new stage in the productivity
literature. Earlier studies primarily asked how productivity should be
\emph{measured}; these systems attempt to improve productivity by changing
the programming and compilation environment itself. They demonstrate that
architecture-specific complexity can be hidden, transformed, or deferred
to compiler and runtime infrastructure.

From a measurement perspective, however, the distinction remains:

\begin{equation}
\boxed{
\text{engineering support for productivity}
\neq
\text{quantified productivity measure}
}
\label{eq:engineering_not_metric}
\end{equation}

This distinction does not diminish the engineering contribution of these
systems. Rather, it clarifies what their evaluations establish. Performance
can be benchmarked, device coverage can be demonstrated, and code size or
programming-interface complexity can be compared. The amount of human
effort avoided, the learning cost introduced, and the maintenance burden
over the software life cycle generally remain separate questions.


\section{Automated Migration and AI-Assisted Programming}
\label{sec:automation_ai}

The increasing automation of software development introduces a challenge that
was largely absent from the early HPC productivity literature. Historically,
source-code size and source-code change could at least be interpreted as rough
proxies for programming work because most transformations were performed
directly by programmers. Source-to-source migration tools and generative AI
weaken this correspondence.

\subsection{Automatic Programming-Model Migration}
\label{subsec:auto_migration}

Tools such as the Intel DPC++ Compatibility Tool and the open-source SYCLomatic
project assist in transforming CUDA applications into SYCL
\cite{IntelDPCT,SYCLomatic}. Such tools can rewrite language constructs and
library calls across substantial portions of a codebase. The generated result
still requires developer review, correction of unsupported transformations,
validation, and often performance tuning, but the number of automatically
modified source lines is no longer a reasonable approximation of the amount of
manual editing performed.

Automation need not take the form of explicit source-to-source migration.
JLIR provides a complementary compiler-level example
\cite{Miniskar2026JLIR}. Its transformation pipeline can identify parallel
and reduction patterns in ordinary serial Julia code and automatically
lower them to portable JACC accelerator kernels without user annotations.
The resulting accelerator implementation may therefore differ
substantially from the programmer-visible source even though the developer
did not manually perform the corresponding transformation.

This exposes an important distinction between at least three quantities:
\emph{transformation magnitude}, \emph{human development effort}, and
\emph{quality of the resulting implementation}. An automatic tool may produce a
large source-level or semantic difference with limited manual intervention.
Conversely, a small residual portion that cannot be migrated automatically may
require substantial architectural knowledge and debugging effort.
Therefore,
\begin{equation}
    \text{source-transformation magnitude}
    \not\propto
    \text{human development effort}.
\label{eq:transformation_not_effort}
\end{equation}
This does not make SLOC, code divergence, or TBMD useless. They continue to
describe properties of the resulting code transformation. It does, however,
weaken their interpretation as productivity measures when automation performs a
substantial fraction of that transformation.

Automatic migration also complicates time-based productivity. The relevant
human effort may no longer be dominated by writing the translated code, but by
preparing the source for migration, interpreting diagnostics, validating
correctness, analyzing performance regressions, replacing unsupported
constructs, and tuning the generated implementation. A meaningful productivity
study must therefore specify which stages of the migration workflow are
included in the measured effort.

\subsection{Generative AI and the Changing Meaning of Programming Effort}
\label{subsec:generative_ai}

Generative AI broadens the automation problem beyond source-to-source
migration. AI-assisted development tools can generate, translate, explain,
debug, restructure, and optimize source code, potentially changing both the
amount and the nature of human effort required to produce a software artifact.
Recent software-engineering studies report measurable productivity gains from
AI assistance, but they also show that the magnitude of those gains depends
strongly on what is chosen as the productivity measure. Peng et
al.~\cite{Peng2023}, for example, reported substantially faster completion of
a programming task with GitHub Copilot, while later studies with professional
developers have also observed reductions in development time
\cite{Paradis2025}.

The difficulty is particularly acute in HPC. Teranishi et
al.~\cite{Teranishi2026AIHPC} identify productivity and performance
portability as two major targets for AI-assisted HPC software development.
They emphasize that general code and text corpora do not directly encode
important runtime properties such as parallel performance, portability,
and scientific reproducibility. Consequently, successful code generation
does not by itself establish that an AI-assisted implementation satisfies
the properties that determine its usefulness in HPC.

More importantly for the present review, recent results demonstrate that
different observables can produce substantially different assessments of the
same AI-assisted development process. Gambacorta et
al.~\cite{Gambacorta2026}, for example, reported a considerably larger
productivity improvement when measured through code output than when measured
through completed tasks. This difference directly illustrates why the amount of
source code produced cannot by itself be identified with programmer
productivity. AI can increase the volume of generated code much more than it
increases useful completed work.

AI assistance can also affect dimensions that are not represented by either
code volume or development speed. Empirical studies have reported incorrect or
non-reproducible generated solutions~\cite{Dakhel2023}, security concerns under
some forms of AI assistance~\cite{Perry2023}, and reduced developer
understanding even when task completion improves~\cite{MartinLopez2026}.
Consequently, faster code generation or shorter task-completion time does not
necessarily imply a proportional improvement in the overall software
development process.

These observations are especially important for the distinction between
artifact and process measures developed throughout this review. SLOC, code
churn, source divergence, and semantic divergence describe properties of the
resulting software artifact. When substantial portions of that artifact are
generated automatically, these measures may record a large transformation even
when comparatively little direct human editing was required. Development time
is closer to a process measure, but AI may change what that time represents:
human effort can shift from writing code toward specifying the task, evaluating
generated solutions, correcting errors, verifying correctness, assessing
security, analyzing performance, and iteratively refining the result.

This shift is particularly consequential in HPC. A generated implementation may
be syntactically correct and functionally valid while still failing to exploit
the target architecture effectively or delivering unacceptable performance.
Thus, neither the amount of generated code nor successful compilation is a
sufficient development outcome for HPC productivity assessment.

Generative AI therefore strengthens the need to distinguish three different
questions:
\begin{enumerate}
    \item \emph{What changed?} --- the magnitude and structure of the resulting
    software transformation;
    \item \emph{How was it produced?} --- the amount and nature of human and
    automated effort involved; and
    \item \emph{What was achieved?} --- correctness, performance, portability,
    maintainability, and other relevant outcomes.
\end{enumerate}
A productivity measure that observes only the first of these cannot reliably
characterize the second, and a measure that captures development effort without
considering the third does not establish whether that effort produced a useful
computational result.


\section{Comparative Synthesis of HPC Productivity Methods}
\label{sec:synthesis}

The historical progression reviewed above should not be interpreted as a simple
sequence in which each new metric replaces the previous one. Instead, the
methods measure different aspects of software development and answer different
questions.

SLOC, LLOC, Halstead complexity, cyclomatic complexity, code divergence, and
TBMD are primarily \emph{artifact measures}. Their major advantage is
reproducibility: once the source code and measurement rules are available,
another researcher can usually recompute the result without reconstructing the
original development process. Their principal limitation is that the inference
from artifact complexity to human effort must be assumed rather than directly
observed.

Development time and developer diaries are \emph{process measures}. They are
closer to the effort that productivity is usually intended to represent, and
they can capture debugging, tuning, learning, and other work invisible in the
final source. Their disadvantage is that the original process must be observed
while it occurs. The measurements are difficult to reconstruct retrospectively
and are strongly affected by programmer expertise and experimental conditions.

Subjective quantities such as perceived difficulty or frustration constitute a
third category. These measures can be important in usability studies because
they capture aspects of programmer experience that objective source metrics
cannot observe. However, they are particularly difficult to use as general
quantitative characteristics of a programming model because they depend on
prior knowledge, expectations, familiarity, and individual perception.

Performance-aware approaches such as RDTP and the Wienke performance-life-cycle
model address another weakness of pure effort metrics: effort alone does not
indicate whether a useful computational result was obtained. These methods
therefore move closer to the conventional economic meaning of productivity as
output divided by input. Their difficulty is that both the effort term and the
functional relationship between effort and performance must be defined and
measured.

Finally, code divergence, code convergence, and TBMD represent attempts to
connect productivity with portability and performance portability. They improve
the characterization of specialization and maintenance burden and permit useful
joint visualization with performance-portability metrics. However, they remain
predominantly artifact-based proxies.

Recent language and compiler systems form a different category. JACC,
JACC.Multi, Mojo, JLIR, and Accl++ are not productivity metrics; they are
\emph{engineering mechanisms intended to improve productivity}. They move
architecture-specific work into common programming abstractions,
intermediate representations, compiler transformations, or runtime
compilation. Their evaluations show that such mechanisms can preserve
competitive performance while reducing some forms of platform-specific
programming. Nevertheless, the resulting productivity claims are usually
supported through indirect observables---common-source programming, fewer
annotations, ecosystem reuse, source-code size, or reduced implementation
duplication---rather than through direct measurement of human effort.

This distinction adds another dimension to the historical progression:
a system may be designed successfully \emph{for} productivity without
providing a metric \emph{of} productivity. The former is an engineering
property of the programming environment; the latter is a measurement
problem.

The central distinction that emerges from the literature can therefore be
expressed as
\begin{equation}
    \underbrace{\text{measured observable}}_{\text{SLOC, time, divergence, TED, \ldots}}
    \quad\neq\quad
    \underbrace{\text{productivity construct}}_{\text{what the observable is intended to represent}}
\label{eq:observable_construct}
\end{equation}
A metric can be objective and reproducible while still being an imperfect proxy
for the construct of interest. Conversely, a direct measure of human effort can
have stronger conceptual validity while being difficult to reproduce or compare
across programmers.

This trade-off explains why no single method has displaced the others. The
simplest measures are attractive precisely because they are inexpensive and
repeatable, while the richer measures require increasingly intrusive data
collection or sophisticated analysis infrastructure.


\section{Open Challenges for Quantified HPC Productivity}
\label{sec:open_challenges}

\subsection{A Workable Formal Definition}
\label{subsec:challenge_definition}

The most basic open problem is the absence of a generally accepted formal
definition that specifies what quantity HPC productivity is intended to measure.
The generic output/input definition is conceptually attractive, but both
``output'' and ``input'' can represent many different quantities. Output may
mean correctness, scientific throughput, speedup, efficiency, or performance
portability; input may mean development time, monetary cost, human effort,
source modification, or maintenance burden.

Without a specified numerator and denominator, apparently conflicting
productivity results may simply correspond to different definitions. A useful
formalization must therefore identify not only a mathematical expression but
also the semantics and units of the quantities entering that expression.

The recent compiler and language systems reviewed in
Section~\ref{sec:productive_compilers} make this requirement especially
visible. A programming environment can credibly reduce architecture-specific
coding, reuse a common ecosystem, or automate accelerator generation without
thereby defining a numerical productivity quantity. Engineering mechanisms
that facilitate productive programming must therefore be distinguished from
the observable chosen to evaluate the productivity that they are intended to
improve.

\subsection{Separating Artifact Change from Human Effort}
\label{subsec:challenge_effort}

A second challenge is construct validity. Code size, churn, divergence, and
tree-edit distance are properties of software artifacts. Human development
effort is a property of the process that produced them. Correlation between the
two may exist in particular settings, but an identity should not be assumed.

This problem is increasingly important as migration tools and AI-generated code
become common. A productivity methodology intended to remain meaningful in
modern software development must either measure human involvement directly or
make explicit that it measures artifact transformation rather than human effort.

\subsection{Relating Effort to Performance}
\label{subsec:challenge_performance}

Greater development effort does not guarantee greater performance. A long
optimization attempt may produce little improvement, whereas a small change can
occasionally produce a large gain. Failed approaches can consume significant
effort without appearing in the final code at all. Conversely, highly optimized
code may embody years of accumulated expertise that cannot be inferred from the
size of the final implementation.

The relationship is therefore neither necessarily linear nor monotonic.
Performance-life-cycle models recognize this explicitly, but a generally
applicable relationship has not been established. Any quantitative
performance--productivity formulation must consequently distinguish the
\emph{amount of effort invested} from the \emph{performance consequence of that
effort}.

\subsection{Connecting Productivity to Portability}
\label{subsec:challenge_portability}

This suggests that portability may influence productivity indirectly.
When an implementation loses portability across the target platform set,
the developer may face increased \emph{adaptation pressure}, by which we
mean the need for platform-specific porting, specialization, or tuning.
The hypothesized relationship can therefore be summarized as
\begin{equation}
\begin{aligned}
    \text{loss of portability}
    &\longrightarrow \text{adaptation pressure} \\
    &\longrightarrow \text{potential additional development effort}.
\end{aligned}
\label{eq:portability_effort_chain}
\end{equation}

However, this relationship remains a hypothesis rather than an established
quantitative law. A portable implementation can perform poorly everywhere, and
a highly specialized implementation can be inexpensive to create if generated
automatically. Portability therefore cannot simply be equated with
productivity.

The reviewed PP--CC and performance-portability--TBMD approaches provide useful
joint views of the dimensions, but the literature has not yet demonstrated a
general mathematical relationship by which a productivity quantity follows from
a performance-portability quantity, or vice versa.

\subsection{Automation and AI}
\label{subsec:challenge_ai}

Automation creates a new requirement for productivity measures: the result
should remain interpretable when humans do not directly write every transformed
line of source code. Measuring generated lines, changed lines, or semantic
distance may continue to characterize software transformation, but these
quantities cannot automatically be interpreted as human input.

Future studies will therefore need to report the development workflow in
addition to the resulting artifact: which transformations were automatic,
which required human intervention, how verification was performed, and what
additional tuning was necessary to recover correctness and performance.

A further difficulty is that AI-assisted development may improve one
productivity observable while degrading another. Faster task completion, more
generated code, and greater feature throughput do not necessarily imply better
correctness, security, maintainability, or developer understanding. Recent
empirical studies illustrate precisely this multidimensional behavior
\cite{Dakhel2023,Perry2023,MartinLopez2026,Gambacorta2026}. A future productivity
metric should therefore specify what constitutes useful output and should avoid
treating code production itself as the output of interest. In particular,
verification and correction effort cannot be ignored merely because the initial
source code was generated automatically.

\subsection{Baselines, Comparability, and Reproducibility}
\label{subsec:challenge_baselines}

Most productivity measures are relative to a baseline, whether explicitly or
implicitly. SLOC changes depend on the original source, RDTP depends on the
chosen performance and effort baselines, code divergence depends on the
implementations being compared, and TBMD depends strongly on the reference
codebase.

The choice of baseline therefore forms part of the metric semantics. A serial
baseline, for example, may be natural for an experiment on parallelization but
less meaningful for a mature application that was designed as parallel
software. Likewise, comparison among programming models is difficult if the
implementations differ in algorithm, maturity, tuning budget, compiler quality,
developer expertise, or use of automated migration.

A practical productivity methodology must therefore combine a formal definition
with sufficiently detailed reporting of the development context to make
cross-study comparison meaningful.

\subsection{Requirements for a Practical Metric}
\label{subsec:challenge_requirements}

The literature reviewed in this paper suggests several requirements for future
quantified productivity methods. A useful measure should have an explicit
definition and interpretation; distinguish direct measurements from proxies;
be objective and reproducible where possible; require sufficiently low
collection overhead for use in real HPC studies; remain meaningful for small
benchmarks as well as larger applications; identify its baseline and
development context; account for the computational outcome rather than effort
alone; and remain interpretable when programming, migration, or optimization is
partly automated.

Equally important, numerical normalization should not be mistaken for a formal
relationship. Two desirable quantities do not become mathematically connected
merely because they have been scaled to the same interval and can therefore be
combined arithmetically.


\section{Conclusions}
\label{sec:conclusions}

The measurement of productivity in high-performance computing has evolved
substantially over the last three decades. Early studies emphasized usability,
programmer behavior, development time, and simple source-code measures. SLOC
became particularly persistent because it is inexpensive, objective, and easy
to reproduce, despite its weak representation of many activities that dominate
parallel software development. More comprehensive approaches subsequently
introduced direct effort logging, software-cost models, and explicit
performance--effort relationships.

The emergence of performance portability shifted the discussion toward the
three Ps of performance, portability, and productivity. Harrell et al.
introduced development and maintenance measures alongside formal
performance-portability analysis. Pennycook et al. used code convergence to
construct a joint performance-portability--productivity navigation space, and
Lin et al. replaced purely textual source divergence with the semantically
richer TBMD approach.

A still more recent stage shifts part of the emphasis from measuring
productivity to engineering programming systems intended to improve it.
JACC and JACC.Multi use common high-level interfaces for heterogeneous
execution, Mojo combines MLIR-based compilation with Python
interoperability, JLIR automates the extraction of portable accelerator
kernels, and Accl++ uses runtime compilation to reduce vendor-specific
programming. These systems demonstrate substantial progress toward
productive performance-portable programming, while also reinforcing the
measurement problem: performance and portability can be evaluated
experimentally, whereas productivity is still commonly inferred through
programming abstraction, ecosystem reuse, code size, or reduced
implementation burden.

Taken together, these developments improve both the measurement of
productivity-related properties and the engineering mechanisms intended to
support productive programming, but they do not eliminate a fundamental
distinction.
Development time measures a process, subjective assessments measure human
experience, and SLOC, divergence, and TBMD measure properties of software
artifacts. None of these quantities can automatically be identified with
productivity in every context.

Moreover, existing attempts to combine productivity proxies with
performance-portability metrics provide useful joint analyses but have not yet
established a general formal relationship between productivity and performance
portability. The rapid growth of automated migration and AI-assisted
programming makes this limitation increasingly important because code quantity
and transformation magnitude are becoming progressively weaker indicators of
direct human effort.

The principal open challenge is therefore not merely to construct another
normalized productivity score. It is to develop a workable formal definition
whose measured quantities have clear semantics, whose relationship to
performance and portability is defensible, and whose interpretation remains
valid as HPC software development becomes increasingly heterogeneous and
automated.


\begin{appendices}
\section{Glossary of Acronyms and Technical Terms}
\label{app:glossary}

For convenient reference, this glossary summarizes the principal acronyms and
specialized terms used throughout the review. The descriptions reflect how the
terms are used in this paper rather than attempting to provide exhaustive
definitions.

\begingroup
\small
\setlength{\tabcolsep}{4pt}
\renewcommand{\arraystretch}{1.10}

\begin{longtable}{@{}
L{0.17\textwidth}
L{0.28\textwidth}
L{0.49\textwidth}
@{}}

\toprule
\textbf{Term} &
\textbf{Expansion or full name} &
\textbf{Meaning in this review} \\
\midrule
\endfirsthead

\multicolumn{3}{@{}l}{\small\textit{Glossary continued}}\\[2pt]
\toprule
\textbf{Term} &
\textbf{Expansion or full name} &
\textbf{Meaning in this review} \\
\midrule
\endhead

\midrule
\multicolumn{3}{r@{}}{\small\textit{Continued on next page}}\\
\endfoot

\bottomrule
\endlastfoot

Accl++ &
--- &
An architecture-agnostic C++-embedded programming approach that uses runtime
compilation to provide code portability across heterogeneous devices. \\

\cmidrule(lr){1-3}

AI &
Artificial Intelligence &
Computer-based generation, transformation, explanation, debugging, or
optimization of source code; the paper focuses particularly on generative
AI-assisted programming. \\

\cmidrule(lr){1-3}

Artifact measure &
--- &
A measure computed from properties of the produced software artifact, such as
source-code size, churn, divergence, or tree structure, rather than from the
development process itself. \\

\cmidrule(lr){1-3}

CC &
Code Convergence &
A quantity defined as one minus Code Divergence; larger values indicate greater
source-code commonality across the considered platform-specific variants. \\

\cmidrule(lr){1-3}

CD &
Code Divergence &
A source-based measure of difference among platform-specific implementations;
in the reviewed formulation it is based on average pairwise source-code
distance. \\

\cmidrule(lr){1-3}

Code churn &
--- &
The amount of source code added, removed, or modified during development. It is
used as an activity or effort proxy rather than as a direct observation of
human effort. \\

\cmidrule(lr){1-3}

COCOMO &
Constructive Cost Model &
A family of software-cost estimation models that predicts development effort
from software size and cost drivers. COCOMO II is the later formulation
discussed in connection with the Wienke HPC effort model. \\

\cmidrule(lr){1-3}

CPU &
Central Processing Unit &
The general-purpose processor of a computing system; used in the paper when
distinguishing CPU and accelerator implementations. \\

\cmidrule(lr){1-3}

CUDA &
Compute Unified Device Architecture &
NVIDIA's GPU programming platform and programming model. It appears both as a
low-level implementation approach and as a source language for automated
migration toward SYCL. \\

\cmidrule(lr){1-3}

Cyclomatic complexity &
--- &
A control-flow complexity measure based on the number of linearly independent
paths through a program's control-flow graph. \\

\cmidrule(lr){1-3}

DPC++ &
Data Parallel C++ &
Intel's C++/SYCL implementation. The Intel DPC++ Compatibility Tool is discussed
as an automated CUDA-to-SYCL migration tool. \\

\cmidrule(lr){1-3}

Function points &
--- &
A software-sizing method based on externally visible functionality such as
inputs, outputs, inquiries, files, and interfaces. \\

\cmidrule(lr){1-3}

GPU &
Graphics Processing Unit &
A highly parallel accelerator processor commonly used in HPC systems. \\

\cmidrule(lr){1-3}

Halstead complexity &
--- &
A family of software-science measures derived from counts of operators and
operands and used to estimate properties such as program volume and effort. \\

\cmidrule(lr){1-3}

HPC &
High-Performance Computing &
Computing focused on achieving high computational performance, typically on
parallel and heterogeneous systems. \\

\cmidrule(lr){1-3}

JACC &
Julia for ACCelerators &
A high-level Julia framework that provides a common front end over multiple
CPU and GPU back ends for performance-portable accelerator programming. \\

\cmidrule(lr){1-3}

Jaccard distance &
--- &
A set-distance measure equal to one minus Jaccard similarity. It is used in the
reviewed Code Divergence formulation to compare sets of source lines. \\

\cmidrule(lr){1-3}

JLIR &
Julia-native Level Intermediate Representation &
A Julia-native, MLIR-inspired intermediate representation framework that can
automatically lower suitable Julia loop structures to portable JACC kernels. \\

\cmidrule(lr){1-3}

LLOC &
Logical Lines of Code &
A source-code size measure based on logical program statements rather than
physical text lines. \\

\cmidrule(lr){1-3}

LOC &
Lines of Code &
A general source-code size measure. Different counting conventions may include
or exclude comments, blank lines, or particular constructs. \\

\cmidrule(lr){1-3}

MLIR &
Multi-Level Intermediate Representation &
An extensible compiler infrastructure that supports multiple abstraction
levels and dialects and enables progressive lowering toward different target
architectures. \\

\cmidrule(lr){1-3}

Mojo &
--- &
An MLIR-based compiled programming language with Python interoperability that
supports vendor-independent GPU programming and is studied here in connection
with performance portability and productivity. \\

\cmidrule(lr){1-3}

MPI &
Message Passing Interface &
A standardized message-passing programming interface widely used for
distributed-memory parallel programming. \\

\cmidrule(lr){1-3}

OpenMP &
Open Multi-Processing &
A directive-based programming interface used for shared-memory parallelism and,
in modern versions, accelerator offload. \\

\cmidrule(lr){1-3}

P3 &
Performance, Portability, and Productivity &
The three dimensions considered together in work on productive performance
portability. \\

\cmidrule(lr){1-3}

Performance portability &
--- &
The ability of an application or implementation to achieve useful performance
or efficiency across a specified set of target platforms. \\

\cmidrule(lr){1-3}

PP &
Performance Portability &
Shorthand used in the paper, figures, and reviewed literature for a quantified
performance-portability measure or score. \\

\cmidrule(lr){1-3}

Process measure &
--- &
A measure of the development process itself, such as development time or logged
developer effort, rather than a property of the final source artifact. \\

\cmidrule(lr){1-3}

Productivity proxy &
--- &
An observable used to infer productivity or development effort without directly
measuring the human input of interest. \\

\cmidrule(lr){1-3}

RDTP &
Relative Development Time Productivity &
A productivity measure that relates achieved speedup to relative development
effort. \\

\cmidrule(lr){1-3}

RTC &
Runtime Compilation &
Compilation performed during program execution or deployment rather than
entirely ahead of time; used by Accl++ to generate device-specific code for
heterogeneous targets. \\

\cmidrule(lr){1-3}

SEER-SEM &
System Evaluation and Estimation of Resources--Software Estimation Model &
A software-cost estimation model cited as an example of a conventional model
that uses software size and project characteristics to estimate effort. \\

\cmidrule(lr){1-3}

SLOC &
Source Lines of Code &
A source-code size measure widely used as an indirect proxy for development
effort and productivity. \\

\cmidrule(lr){1-3}

SLIM &
Software Life-Cycle Management &
A software-cost estimation model cited alongside COCOMO and SEER-SEM. \\

\cmidrule(lr){1-3}

SYCL &
SYCL programming model &
A single-source C++ programming model for heterogeneous systems. In this paper
it appears both as a performance-portable programming approach and as a target
of CUDA migration tools. \\

\cmidrule(lr){1-3}

SYCLomatic &
SYCLomatic migration tool &
An open-source source-to-source migration tool used to assist conversion of CUDA
code to SYCL. \\

\cmidrule(lr){1-3}

TBMD &
Tree-Based Model Divergence &
A codebase-artifact metric that uses compiler-derived tree representations and
Tree Edit Distance to quantify divergence from a selected reference codebase. \\

\cmidrule(lr){1-3}

TED &
Tree Edit Distance &
A distance measure based on the edits required to transform one tree into
another; used by TBMD to compare compiler-derived code representations. \\

\cmidrule(lr){1-3}

UPC &
Unified Parallel C &
A parallel extension of C designed for partitioned global address-space
programming. \\

\cmidrule(lr){1-3}

X10 &
X10 programming language &
A parallel programming language appearing in the earlier empirical productivity
literature reviewed in this paper. \\

\end{longtable}

\endgroup
\end{appendices}

\backmatter


\bibliography{sn-bibliography}

\end{document}